\documentclass{elsart}
\usepackage{amsmath}
\usepackage{amssymb}
\usepackage{latexsym}
\usepackage{epsfig}

\begin{document}
\begin{frontmatter}
\title{On Bose-Einstein condensation on closed Robertson-Walker spacetimes}

\author{M.~Trucks}, 
\address{Institut f\"ur Theoretische Physik\\
Technische Universit\"at Berlin\\
Hardenbergstra\ss e 36, 10623 Berlin, Germany\\
E-mail: trucks@physik.tu-berlin.de}
\date{\today}

\begin{abstract}
  In this letter we summarize our analysis of Bose-Einstein condensation
  on closed Robertson-Walker spacetimes. In a previous work we defined
  an adiabatic KMS state on the Weyl-algebra of the free massive
  Klein-Gordon field \cite{Tru97,Tru98a}. This state describes a
  free Bose gas on Robertson-Walker spacetimes. We use this state to
  analyze the possibility of Bose-Einstein condensation on closed
  Robertson-Walker spacetimes. We take into account the effects due to
  the finiteness of the spatial volume and show that they are not
  relevant in the early universe. Furthermore we show that a critical
  radius can be defined. The condensate disappears above the critical
  radius. 
\end{abstract}
\end{frontmatter}

\section{Introduction}

We analyze the possibility of Bose-Einstein condensation of a free
relativistic charged Bose gas on closed Robertson-Walker spacetimes.
The free Bose gas is described by an adiabatic KMS state, which we
defined in \cite{Tru97,Tru98a}. In the framework of algebraic quantum
field theory on curved spacetimes (see \cite{Wald}), we proved that an
adiabatic KMS state is a Hadamard state, i.e.\   a physically relevant
state. Furthermore, it describes a free Bose gas and the inverse
temperature change was shown to be proportional to the scale parameter
in the metric of the Robertson-Walker spacetime.

Bose-Einstein condensation is usually considered in the thermodynamic
limit. Singularities appear in thermodynamic quantities, e.g.\ for a
nonrelativistic Bose gas, there is a cusp-like singularity in the
specific heat at the critical temperature. For a finite system, these
singularities are smoothed out and there is no well-defined phase
transition point. Furthermore, macroscopic occupation of the ground
state sets in, although the temperature is higher than the critical
temperature of the corresponding system in the thermodynamic limit, as
we will see below.

We can apply known results about Bose-Einstein condensation on the
Einstein universe (a closed Robertson-Walker spacetime with constant
scale parameter) to closed Robertson-Walker spacetimes.  This is
justified on the one hand by the fact that adiabatic KMS states are
Hadamard states and on the other hand because on every Cauchy surface
an adiabatic KMS state on a closed Robertson-Walker spacetime has the
form of a KMS state on an Einstein universe.  Bose-Einstein
condensation on the Einstein universe was analyzed by Singh and
Pathria \cite{SP84} and Parker and Zhang \cite{PZ91}.  In the work of
Singh and Pathria the emphasis is on the finite size effects, while
they are not taken into account in the work of Parker and Zhang. In
contrast the work of Parker and Zhang also gives some hints on
Bose-Einstein condensation on closed Robertson-Walker spacetimes and a
critical radius is defined. We analyze Bose-Einstein condensation on
closed Robertson-Walker spacetimes, take into account the finite size
effects and define a critical radius. If the Robertson-Walker
spacetime expands, there is a radius, below which the gas is in a
condensed state. The condensate disappears above this radius, but due
to the finite size effects, this is not a well-defined phase
transition point. Using realistic values for the early universe, we
will see that the finite size effects are very small.

In the next section we calculate the charge density of the Bose gas on
closed Robertson-Walker spacetimes. An approximation is made for the
interesting regime, where the chemical potential reaches the value of
the mass parameter. This formula is analyzed in
section~\ref{sec:finitesize}. A numerical calculation shows the
effects of the finite spatial volume of closed Robertson-Walker
spacetimes. The critical temperature of a charged relativistic Bose
gas on Minkowski spacetime serves as a useful reference point, which
enables us to define a critical radius. This critical radius depends
only on the total charge and a constant $C=RT$, given by the product
of the temperature and the scale parameter of the Robertson-Walker
metric.  Due to the finite spatial volume, there is a nonvanishing
charge density in the ground state at the critical temperature (resp.\ 
radius). This value (scaled by the charge density) mainly depends on
$C$.

\section{The charge density}

In this section we calculate the charge density of a relativistic
charged Bose gas described by an adiabatic KMS state. An adiabatic
KMS state on closed Robertson-Walker spacetimes can be defined 
on a Cauchy surface at time $t$ by
\begin{eqnarray*}
  \omega(\cdot) = Z^{-1}\mbox{tr}\{\exp[-\beta H(\mu)]\;\cdot\;\},
\end{eqnarray*}
where $H(\mu)$ is the second quantization of $h(\mu) =
[m^{2}-\Delta/R^{2}(t)]^{1/2}-\mu$, $\beta$ is the inverse
temperature, $\mu$ the chemical potential, $R$ the scale parameter in
the metric of the Robertson-Walker spacetime, $Z =
\mbox{tr}\{\exp[-\beta H(\mu)]\}$ and $\Delta$ the Laplace operator on
the three-sphere $S^{3}$, the spatial part of the spacetime. The
conserved charge, which may be the electric or any other conserved
charge, is described by the operator $Q = N_{+} - N_{-}$, where
$N_{+}$ resp.\ $N_{-}$ are number operators for particles with
positive resp.\ negative charge. Then the state is given by
\begin{eqnarray*}
  \omega(\cdot) = Z^{-1}\mbox{tr}\{\exp[-\beta (H-\mu
  Q)]\;\cdot\;\}.
\end{eqnarray*} 
For the expectation value of the charge density in this state we have
\begin{eqnarray*} 
  q = \; <Q>_{\beta,\mu}/V =
  (<N_{+}>_{\beta,\mu}-<N_{-}>_{\beta,\mu})/V. 
\end{eqnarray*}
It is 
\begin{eqnarray*}
  <N_{\pm}>_{\beta,\mu}\ = \sum_{\vec{k}}
  \frac{\e^{\pm\beta\mu}}{\e^{\beta\tilde{\omega}_{k}}-\e^{\pm\beta\mu}},
\end{eqnarray*}
where $\tilde{\omega}_{k} = \sqrt{k(k+2)/R^{2}(t)+m^{2}}$, $k\in
\mathbb{N}_{0}$, since $k(k+2)$ are the eigenvalues of the Laplace
operator on the three-sphere. Using the fact that the multiplicity of
the eigenvalues is $(k+1)^{2}$, we obtain
\begin{eqnarray*}
  q = \frac{2}{V} \sum_{j=1}^{\infty}\sinh(\beta\mu
  j)\sum_{k=1}^{\infty} k^{2}\e^{-\beta\omega_{k}n},
\end{eqnarray*}
where $\omega_{k}=\sqrt{\tilde{m}^{2}+k^{2}/R^{2}}$, $\tilde{m}^{2} =
m^{2}-1/R^{2}$. The sums can be simplified with the Poisson summation
formula 
\begin{eqnarray*}
  \sum_{k=1}^{\infty}f(k) = \int_{0}^{\infty}f(t)dt +
  \sum_{p=1}^{\infty} \int_{0}^{\infty} f(t) \cos(2\pi pt)dt.  
\end{eqnarray*}
The following computation follows \cite{SP84}.  The first integral is
given in \cite[3.389 Nr.4]{GR} and we have to assume $R>1/m$. To solve
the second integral, we expand the cosine in a power series and use
the formula
\begin{eqnarray*}
  \sum_{l=0}^{\infty}\frac{(z\tau^{2}/2)^{l}}{l!}
  \frac{K_{l+\lambda}(zs)}{s^{l+\lambda}} =
  \frac{K_{\lambda}(z\sqrt{s^{2}-\tau^{2}})}
  {(s^{2}-\tau^{2})^{\lambda/2}}  
\end{eqnarray*}
(see \cite[7.15 Nr.9]{EMOT2}), to obtain
\begin{eqnarray*} 
  \int_{0}^{\infty} f(t) \cos(2\pi pt)dt = j\beta\tilde{m}^{4}R^{3}
  \left( \frac{K_{2}(a\tau)}{(a\tau)^{2}} - (2\pi
    p\tilde{m}R)^{2}\frac{K_{3}(a\tau)}{(a\tau)^{3}} \right),
\end{eqnarray*}
where $\tau=\sqrt{(j\beta/R)^{2} + (2\pi p)^{2}}$, $a = \tilde{m}R$
and $K_{\nu}(x)$ are modified Bessel functions. Therefore
\begin{eqnarray*}
  q = \frac{\tilde{m}^{3}}{2\pi^{2}}W(\beta,\mu) +
  \frac{2\beta\tilde{m}^{4}}{\pi^{2}}
  \sum_{j,p=1}^{\infty}j\sinh(j\beta\mu) \left(
    \frac{K_{2}(a\tau)}{(a\tau)^{2}} - (2\pi
    pa)^{2} \frac{K_{3}(a\tau)}{(a\tau)^{3}}
  \right), 
\end{eqnarray*}
where we defined $W(\beta,\mu) =
2\sum_{j=1}^{\infty}(j\beta\tilde{m})^{-1}\sinh(j\beta\mu)
K_{2}(j\beta\tilde{m})$.  We can also handle the sum over $j$ with the
Poisson summation formula in the form
\begin{eqnarray*}
  \sum_{j=1}^{\infty}f(j) &=& \frac{1}{2}\sum_{j=-\infty}^{\infty}f(j)
  = \sum_{p=-\infty}^{\infty}\int_{-\infty}^{\infty}f(j)\cos(2\pi pj)\,dj \\
  &=& \frac{1}{2}\sum_{p=-\infty}^{\infty}\Re
  \int_{0}^{\infty}j\sinh(j\beta\mu') \left(
    \frac{K_{2}(a\tau)}{(a\tau)^{2}} - (2\pi 
    p\tilde{m}R)^{2}\frac{K_{3}(a\tau)}{(a\tau)^{3}} \right)dj
\end{eqnarray*}
see \cite[App.]{SP84} where $\mu' = \mu + 2\pi {\mathrm i}p/\beta$ and
$\Re$ denotes the real part. The result is 
\begin{eqnarray*}
  q = \frac{\tilde{m}}{2\pi^{2}}W(\beta,\mu) - \frac{1}{\pi\beta}\Re
  \left\{ \sum_{l=-\infty}^{\infty}
    \frac{\mu'\sqrt{\tilde{m}^{2}-\mu'{}^{2}}}{\exp(2\pi
      R\sqrt{\tilde{m}^{2}-\mu'{}^{2}})} \right\},
\end{eqnarray*}
The second term comes from the finiteness of the spatial volume of the
system.  It can be shown that all terms with $l \neq 0$ can be
neglected if $R/\beta \gg 1$, an inequality surely satisfied in the
early universe. The term with $l=0$ is the significant term, giving
the main contribution especially if $\mu \to \tilde{m}$.  So we have
\begin{eqnarray*}
  q = \frac{\tilde{m}}{2\pi^{2}}W(\beta,\mu) - \frac{\mu}{\pi^{2}\beta
  R} \frac{\pi R \sqrt{\tilde{m}^{2}-\mu^{2}}}{\exp(2\pi R
  \sqrt{\tilde{m}^{2}-\mu^{2}}) - 1}.
\end{eqnarray*}

The charge density in the ground state is given by 
\begin{eqnarray*}
  q_{0} &=& \frac{1}{2\pi^{2}R^{3}}\left( (\e^{\beta(m-\mu)}-1)^{-1} -
    (\e^{\beta(m+\mu)}-1)^{-1} \right).
\end{eqnarray*}
If $\mu \to m$, we approximate $q_{0}$ by
\begin{eqnarray*}
  q_{0} &\approx& \frac{1}{2\pi^{2}R^{3}}\left( [\beta(m-\mu)]^{-1} -
    [\beta(m+\mu)]^{-1} \right) \\
  &=& \frac{1}{\beta\pi^{2}R^{3}}\frac{\mu}{m^{2}-\mu^{2}} \\
  &=& \frac{\mu}{R\beta(y^{2}+\pi^{2})}, 
\end{eqnarray*}
where we defined the thermogeometric parameter $y = \pi R
\sqrt{\tilde{m}^{2}-\mu^{2}}$ (see \cite{SP84}). The charge density in
the ground state grows rapidly if $y^{2} \to -\pi^{2}$.

If $\mu \to m$ we can expand the function $W(\beta,\mu)$ (see
\cite{SP84,SP83}):
\begin{eqnarray*}
  \frac{\tilde{m}^{3}}{2\pi^{2}}W(\beta,\mu) =
  \frac{\tilde{m}^{3}}{2\pi^{2}} W(\beta,\tilde{m}) -
  \frac{\tilde{m}}{2\pi \beta}\sqrt{\tilde{m}^{2}-\mu^{2}} +
  O(\tilde{m}^{2}-\mu^{2}), 
\end{eqnarray*} 
which leads to 
\begin{eqnarray}\label{chargedensity}
  q \approx \frac{\tilde{m}^{3}}{2\pi^{2}}W(\beta,\tilde{m}) -
  \frac{\tilde{m}}{2\pi^{2}R\beta}y\coth y \approx
  \frac{m^{3}}{2\pi^{2}}W(\beta,m) - \frac{m}{2\pi^{2}R\beta} \;
  y\coth y \,.  
\end{eqnarray}

\section{The finite size effects}
\label{sec:finitesize}
We have to solve formula (\ref{chargedensity}) for the unknown value
$\mu$. This can only be done numerically. It is well-known that for
finite systems there is no well-defined critical temperature. The critical
temperature of the system in the thermodynamic limit can only serve as a
useful reference point. For the system in the thermodynamic limit,
i.e.\ for Minkowski spacetime, the critical (inverse) temperature is
given by 
\begin{eqnarray*}
  q = q(\beta_{c},m) = \frac{m^{3}}{2\pi^{2}}W(\beta_{c},m).
\end{eqnarray*}
Using this reference point, we get as an equation for $y$ and also for
$\mu$, 
\begin{equation}\label{eq:ycothy}
  y\coth y = -\frac{2\pi^{2}Rq\beta}{m} \left( 1 -
    \frac{W(\beta_{c},m)}{W(\beta,m)}\right),
\end{equation}
which can be used to determine the charge density in the ground state
$q_{0}$.  We have used this equation to analyze the finite size
effects. This is shown in fig.~1. It can be seen clearly that there is
no well-defined phase transition point. Enlarging the radius $R$
forces the curve for the Einstein universe towards the curve for
Minkowski spacetime.
\begin{figure}
\begin{center}
\epsfig{file=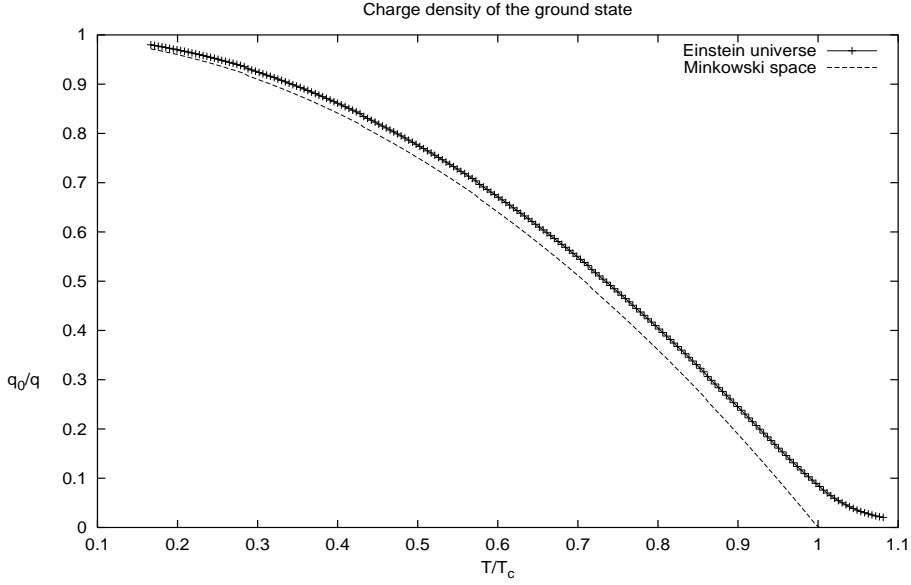,height=8cm,width=12cm}\\[0.5cm]
\caption{Charge density of the ground state $q_{0}$ for a closed 
  Robertson-Walker spacetime on a Cauchy surface, where $R = 10$
  (thick line) compared with the system in the thermodynamic limit,
  i.e.\ Minkowski spacetime (thin line) and $Q=0.01$. }
\end{center}
\end{figure}

The critical temperature $T_{c}$ of Minkowski spacetime, given by
\begin{eqnarray*}
  T_{c} = \sqrt{3|q|/m}
\end{eqnarray*}
(see Haber and Weldon \cite{HW81}), serves as a useful reference point.
Using the fact that $RT =: C = \mbox{const.}$ (proved in
\cite{Tru97,Tru98a}), we can define a critical radius $R_{c}$,
instead of the critical temperature, given by
\begin{equation}
  \label{eq:R-c}
  R_{c} = \frac{3|Q|}{2\pi^{2}mC^{2}},
\end{equation}
where $Q=qV=q2\pi^{2}R^{3}$ is the total charge. This was already
realized by Parker and Zhang \cite{PZ91}. Below this radius, the gas is
in a condensed state. If the radius becomes larger than the critical
radius, the condensate disappears.

The question arises, whether, due to the finite size effects, there is
a considerable charge density of the ground state at the critical
radius and whether the lifetime of the condensate is considerably
longer than indicated by $R_{c}$.

For $T=T_{c}$, we have $y\coth y = 0$ (see eq.\ (\ref{eq:ycothy})),
i.e.\ $y^{2}=-\pi^{2}$, and since $y=\pi R\sqrt{\tilde{m}^{2} -
  \mu^{2}}$, $\mu=\sqrt{m^{2}-3/(4R_{c}^{2})}$.  This leads to
\begin{eqnarray*}
  \frac{|q_{0}|}{|q|} &=& \frac{1}{|q|}\frac{T_{c}}{R_{c}}
  \frac{\mu}{\pi^{2}+y^{2}} = \sqrt{\frac{3}{m|q|R_{c}^{2}}}
  \frac{\sqrt{m^{2}-3/(4R_{c}^{2})}}{3\pi^{2}/4} = \\
  &=& \sqrt{\frac{3V}{m|Q|R_{c}^{2}}} \frac{4}{\pi^{2}}
  \sqrt{m^{2}-3/(4R_{c}^{2})} = \\
  &=& \sqrt{\frac{2\pi^{2}R_{c}}{3m|Q|}}\sqrt{m^{2}-3/(4R_{c}^{2})}=\\
  &=& \frac{4}{m\pi^{2}C} \sqrt{m^{2}-3/(4R_{c}^{2})} \approx \\
  &\approx & \frac{4}{\pi^{2}C}.
\end{eqnarray*}
We see that the charge density in the ground state mainly depends on
$C$, the product of $R$ and $T$, which is constant as long as the gas
can be considered as relativistic. Since this is a large value in the
early universe, the finite size effects do not play a considerable
role. 

The critical radius defined in eq.\ (\ref{eq:R-c}), below which the
Bose gas is in a condensed state, makes it possible to think of the
following scenario. If there is an abundance of one charge (not
necessarily the electric charge) in the early universe, the Bose gas
is in a condensed state. The condensate disappears if the radius is
larger than the critical radius and the Bose gas is in its usual
state. We are not aware of any consequences, a condensed Bose gas
would have for the early evolution of the universe, but we think it is
worth investigating such consequences.

We give an approximative value for the electric charge (see also
\cite{PZ91}).  Results of redshift observations \cite{Seg90} show that
the present scale factor $R_{0} \geq 10^{27}\;\mbox{cm} \approx
10^{41}\mbox{GeV}^{-1}$ and results for an upper bound on the present
charge density \cite{DZ81} give $q_p < 10^{-24}\; \mbox{cm}^{-3}
\approx 10^{-65}\;\mbox{GeV}^3$.  Using a value $C\approx10^{28}$ and
$m = m_{\pi}$, the pion mass, we find
\begin{eqnarray*} \label{eq:1}
  R_c \approx 10^5 \;\mbox{GeV}^{-1} \approx 10^{-9} \;\mbox{cm} \,.
\end{eqnarray*}
We haven taken the maximal value of the electric charge density and
the minimal value for the present scale parameter. So the value may
differ in both directions considerably. This may be taken as a
challenge to find more accurate bounds, especially for the present
charge density. In this context it may also be interesting to find
bounds for the charge density of other charges.  Nevertheless even if
the universe is neutral with respect to all charges in question, there
may be regions with positive charge density as well as other regions
with a negative one, so that condensation in special regions occurs.

\section{Conclusions}

We analyzed the possibility of Bose-Einstein condensation of a free
relativistic charged Bose gas on closed Robertson-Walker spacetimes,
described by an adiabatic KMS state.  Macroscopic occupation of the
ground state sets in, if the temperature is in the region of the
critical temperature of a charged relativistic Bose gas on Minkowski
spacetime.  We have shown the effects due to the finiteness of the
spatial volume of closed Robertson-Walker spacetimes but also that
they are not relevant in the early universe. Furthermore a critical
radius can be defined, above which the condensate disappears. This
critical radius depends only on the total charge and the constant
product of the temperature and the scale parameter in the
Robertson-Walker metric. We also pointed out that this critical radius
may have some relevance in cosmology. A next step could be to consider
a self-interacting charged scalar field. From a cosmological point of
view it may also be interesting to consider the consequences of a
condensed Bose gas for the evolution of the universe. Furthermore
there is a challenge to find more accurate bounds on the charge
density of charged bosons.


\begin{thebibliography}{10}
  
\bibitem{Tru97} M.~Trucks, M.~Keyl, The free {B}ose gas on
  {R}obertson-{W}alker spacetimes, Phys.\ Lett.\ B 399 (1997)
  223--226.
  
\bibitem{Tru98a} M.~Trucks, {A} {KMS}-like state of {H}adamard type on
  {R}obertson-{W}alker spacetimes and its time evolution, Commun.\ 
  Math.\ Phys.\ 197 (1998) 387--404, gr-qc/9709019.
  
\bibitem{Wald} R.~Wald, {Q}uantum {F}ield {T}heory in {C}urved
  {S}pacetime and {B}lack {H}ole {T}hermodynamics (University of
  Chicago Press, 1994).
  
\bibitem{SP84} S.~Singh, R.~Pathria, {B}ose-{E}instein condensation in
  an {E}instein universe, J.\ Phys.\ A: Math.\ Gen.\ 17 (1984)
  2982--2994.
  
\bibitem{PZ91} L.~Parker, Y.~Zhang, {U}ltrarelativistic
  {B}ose-{E}instein condensation in the {E}instein universe and energy
  conditions, Phys.\ Rev.\ D 44 (1991) 2421--2431.
  
\bibitem{GR} I.~Gradshteyn, I.~Ryzhik, {T}able of {I}ntegrals,
  {S}eries and {P}roducts (Academic Press, 1980).
  
\bibitem{EMOT2} A.~Erdeleyi et~al., {H}igher {T}ranscendental
  {F}unctions {V}ol.\ 2 (Mc-Graw Hill, 1953).
  
\bibitem{SP83} S.~Singh, P.~Pandita, {S}caling and universality of
  thermodynamics and correlations of an ideal relativistic {B}ose gas
  with pair production, Phys.\ Rev.\ A 28 (1983) 1752--1761.
  
\bibitem{HW81} H.~Haber, H.~Weldon, {T}hermodynamics of an
  ultrarelativistic {B}ose gas, Phys.\ Rev.\ Lett.\  46 (1981)
  1497--1500.
  
\bibitem{Seg90} I.~Segal, Mon.\ N.\ Roy.\ Astron.\ Soc.\  242 (1990)
  423.
  
\bibitem{DZ81} A.~Dolgov, Y.~Zeldovich, Cosmology and elementary
  particles, Rev.\ Mod.\ Phys.\   53 (1981) 1--41.

\end{thebibliography}

\end{document}